\input psfig.tex
\documentclass{aa}
\usepackage{latexsym}
\begin{document}
\thesaurus{ 01		% A&A Section 1: Letters 
          (02.16.2; 	% Polarization       
           09.13.1; 	% ISM: magnetic fields 
           09.19.1; 	% ISM: structure            
           13.18.3)} 	% Radio continuum: ISM          
                     
\title{Structure in the local Galactic ISM on scales down to 1 pc, 
       from multi-band radio polarization observations}

%\subtitle{Structure in the local Galactic ISM}
\author{M.~Haverkorn \inst{1}, P.~Katgert \inst{1}, 
        A.G.~de~Bruyn\inst{2,3} 
}
\institute{
           Sterrewacht Leiden, P.O. Box 9513, 2300 RA Leiden, The
	   Netherlands (haverkrn,katgert@strw.leidenuniv.nl) \and
	   ASTRON, P.O. Box 2, 7990 AA Dwingeloo, The Netherlands
	   (ger@nfra.nl) \and 
	   Kapteyn Institute, P.O. Box 800, 9700 AV Groningen, The 
	   Netherlands }

\offprints{haverkrn@strw.leidenuniv.nl}
\date{Received date; accepted date}

\titlerunning{Structure in the local Galactic ISM on scales down to 1
pc}
\maketitle
%\markboth{}

\begin{abstract}

We discuss observations of the linearly polarized component of the
diffuse galactic radio background. These observations, with an angular
resolution of 4$^{\prime}$, were made with the Westerbork Synthesis
Radio Telescope (WSRT) in 5 frequency bands in the range 341 -- 375
MHz. The linearly polarized intensity $P$ (with polarized brightness temperature %$T_{\rm b,pol}$ 
going
up to 10~K) shows a `cloudy' structure, with characteristic scales of
15 -- 30$^{\prime}$, which contains relatively long, but very narrow
`canals' (essentially unresolved) in which $P$ is only a small
fraction of that in the neighbouring beams.

These `canals' are generally seen in more than one frequency band,
although their appearance changes between bands. They are probably due
to depolarization within the synthesized beam, because the change in polarization angle 
$\Delta\phi_{\rm pol}$ across the deepest `canals' is in general close
to $90^{\circ}$ (or $270^{\circ}$ etc.). These very abrupt changes in
%polarization angle
$\phi_{\rm pol}$, which are seen only across the `canals', seem to
be accompanied by abrupt changes in the Rotation Measure (RM), which
may have the right magnitude to create the difference of close to
$90^{\circ}$ in 
%polarization angle
$\phi_{\rm pol}$, and thereby the `canals'.

The structure in the polarization maps is most likely due to Faraday
rotation modulation of the probably smooth polarized radiation emitted
in the halo of our Galaxy by the fairly local ($\la 500$ pc)
ISM. Therefore, the abrupt changes of RM across the `canals' provide
evidence for very thin ($\la 1$ pc), and relatively long transition
regions in the ISM, across which the RM changes by as much as
100\%. Such drastic RM changes may well be due primarily to abrupt
changes in the magnetic field.

\end{abstract}


\section{Introduction}
\label{s-intro}

Wieringa et al.\ (1993) were the first to note structure on arcminute
scales in the linearly polarized component of the galactic radio
background at 325 MHz, observed with the WSRT.\@ The small-scale
structure in the maps of polarized intensity $P$, (with polarized
brightness temperatures $T_{\rm b,pol}$ of up to 10~K) does NOT have a
counterpart in total intensity, or Stokes $I$, down to very low
limits.  Because the total Stokes $I$ of the galactic radio background
has an estimated $T_{\rm b,pol}$ of the order of 30 -- 50 K at
325 MHz, which must be very smooth and therefore filtered out
completely in the WSRT measurements, the {\em apparent} polarization
percentage of the small-scale features can become very much larger
than 100\%.

The absence of corresponding small-scale structure in Stokes $I$ led
Wieringa et al.\ (ibid.) to propose that the small-scale structure in
polarized intensity $P$ is due to Faraday rotation modulation. In this
picture, synchrotron radiation generated in the Galactic halo reaches
us through a magneto-ionic screen, viz. the warm relatively nearby
ISM. Structure in the electron density and/or magnetic field in the
ISM causes spatial variations in the Rotation Measure (RM) of the
screen.  Hence, the angle of linear polarization of the synchrotron
emission from the halo is rotated by different amounts along different
lines of sight. Even if the polarized emission in the halo were
totally smooth, in intensity as well as angle, the screen would
produce structure in Stokes $Q$ and $U$. 

Small-scale structure in the polarized galactic radio background
has recently been observed also at other frequencies.  At 1420 MHz,
Gray et al.\ (1998, 1999) used the DRAO synthesis telescope to study
the phenomenon at 1$^{\prime}$ resolution. Uyaniker et al.\ (1999)
used the Effelsberg telescope at 1.4 GHz, to map the polarized
emission at 9$^{\prime}$ resolution over about 1100$^\Box$. Duncan et al.\ (1998) discuss radio polarization data at 1.4,
2.4 and 4.8 GHz with the Parkes radio telescope and the VLA, at resp.\
5$^{\prime}$, 10$^{\prime}$ and 15$^{\prime}$ resolution. All these
observations support the interpretation in terms of modulation of
emission originating at larger distances, by a relatively nearby
Faraday screen.

The distributions of polarized intensity and angle may therefore be
used to study the structure of the Faraday screen. In particular,
polarization observations give information about the electron density,
n$_{\rm e}$, and the component of the magnetic field parallel to the
line of sight, $B_{\|}$, in the ISM on scales down to less than $\sim$
0.5 pc ($<$ 4$^{\prime}$ at an assumed distance of $\sim$ 500 pc). The
diffuse nature of the polarized radio background allows (almost)
complete spatial mapping of RMs over large areas, provided one has
observations at several frequencies. This gives a large advantage over
RM determinations through individual objects, like pulsars or
extra-galactic radio sources.

\section{Distribution of polarized intensity} 
\label{s-polint}

\begin{figure}
\vbox{}
\psfig{file=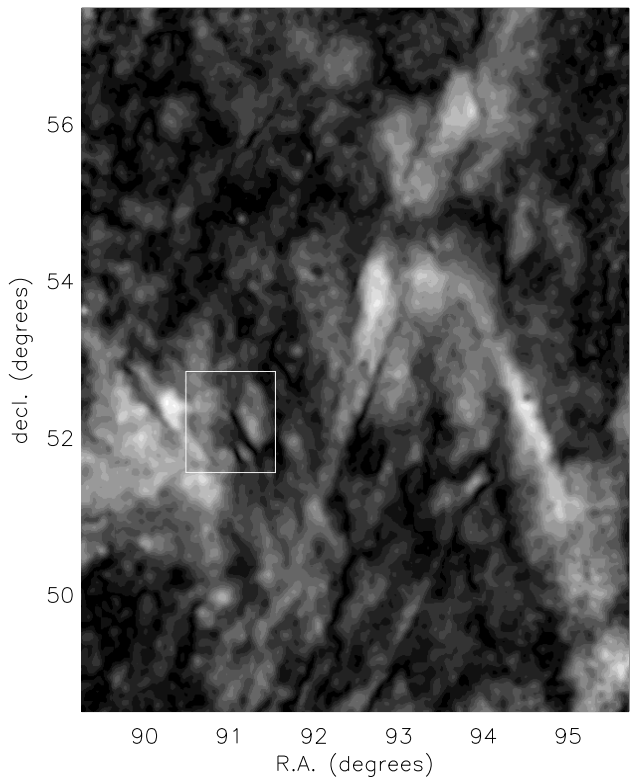,width=8.5cm}
\caption[]{Linearly polarized intensity $P$ at 349~MHz in a 
           $6.4^{\circ}\times9^{\circ}$ field centered at
           $\ell = 161^{\circ}, b = 16^{\circ}$. The resolution is
           $\sim$ 4$^{\prime}$, the maximum brightness temperature is
           $\sim$ 10~K. The generally `cloudy' distribution contains
           long narrow `canals' of low $P$.\@ The white box shows the area
           displayed in Fig.~\ref{f-phi349}.}
\label{f-pol349}
\end{figure}

In Fig.~\ref{f-pol349} we show a gray-scale representation of the
polarized intensity in a 5~MHz wide frequency band centered at
349~MHz. The map shows a region of $6.4^{\circ}\times9^{\circ}$
centered at $\alpha = 6^h10^m,
\delta = 53^{\circ} (\ell = 161^{\circ}, b = 16^{\circ})$ at an
angular resolution of about 4$^{\prime}$. It is one of 8 frequency
bands observed simultaneously. Three of those have strong
interference, but we obtained good data at 341, 349, 355, 360 and 375
MHz. All 5 maps were made combining mosaics of 7$\times$5 pointing
centres.  This yields constant sensitivity over a large area (see e.g.\
Rengelink et al.\ 1997). The observations were made with the WSRT in
January and February 1996, largely at night, and ionospheric Faraday
rotation was therefore well-behaved. No corrections were applied.

The region in Fig.~\ref{f-pol349} is rather special because $T_{\rm
b,pol}$ goes up to 10~K, and because it contains large, almost linear
structures in $P$.\@ Our attention was drawn to this field by the
panoramic view of galactic polarization produced in the WENSS
survey (de Bruyn \& Katgert 2000).  However, this field is not unique, and
there are other regions with similarly high $T_{\rm b,pol}$. Over a
very large fraction of the map the $P$-signal is quite significant, with a noise $\sigma_{\rm T_b} \approx$ 0.5~K.\@ With S/N-ratios of
generally more than 3 and going up to 30, polarization angles are
well-defined. Note that in this region, the upper limit to structure in
Stokes $I$ (total intensity) on small scales ($\la$ 30$^{\prime}$) is
about 1~K, or less than 2\% of the total $I$.

%The structure of the polarized intensity distribution cannot easily be
%summarized in words (or numbers). However, 
There appear to be at least
two distinct components in the polarized intensity distribution. The
first one is a fairly smooth, `cloudy' component, pervading the entire
map, with intensity variations on typical scales of (several) tens of
arcminutes. In addition, there are conspicuous, very narrow and often
quite long and wiggly structures, which we will refer to as `canals',
in which the polarized intensity is considerably lower than in the
immediate surroundings. In this Letter we focus on the nature and
implications of the narrow `canals'; we will discuss the `cloudy'
component in more detail in another paper (Haverkorn et al.\ 2000).

\section{The nature of the `canals' in polarized intensity}
\label{s-canals}

The strong and abrupt decrease of polarized intensity in the `canals'
suggests that depolarization is responsible. There are several
mechanisms that can produce depolarization, but the only plausible
type in this case is beam depolarization. This occurs when the
polarization angle varies significantly within a beam. Complete
depolarization requires that for each line of sight there is a
`companion' line of sight within the same beam that has the same
polarized intensity but for which the polarization angle differs by
90$^{\circ}$. Below we will show that our observations indicate that
the polarization angle indeed changes by large amounts across low
polarized intensity `canals', and close to 90$^{\circ}$ across the
`canals' of lowest $P$.

Depolarization can also be caused by `differential Faraday rotation'.
This happens when along a line of sight emitting and (Faraday)
rotating plasmas coexist (e.g.\ Burn 1966; Sokoloff et al.\ 1998).
However, the absence of correlated structure in Stokes $I$ and the
high degree of polarization suggest that this is not a dominating
effect. Significant bandwidth depolarization, which occurs when the
polarization angle is rotated by greatly different amounts in
different parts of a frequency band could only play a r\^ole (given
our 5~MHz bandwidth) if the RM were of order 80 rad m$^{-2}$, which is
not the case in this region near the galactic anti-centre (see below).

\begin{figure}
\vbox{}
\psfig{file=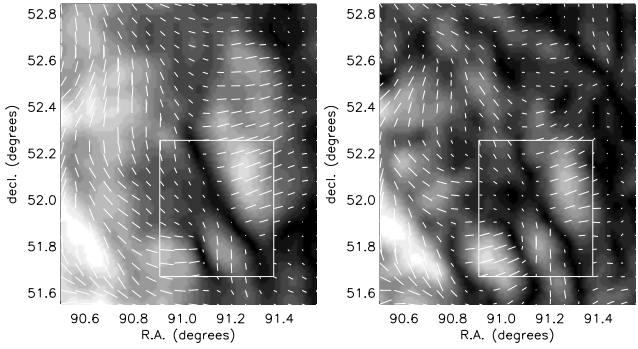,width=8.5cm}
\caption[]{Polarized intensity $P$ at 349~MHz (left) and 360~MHz (right) 
           of the area inside the box in Fig.~\ref{f-pol349}.
           Polarization angles and intensities are indicated by the
           vectors, which are sampled at locations 4$^{\prime}$ apart
           (independent beams). Note $\Delta\phi_{\rm pol} \approx
           90^{\circ}$ across low-$P$ `canals'.}
\label{f-phi349}
\end{figure}

In Fig.~\ref{f-phi349} we show the polarization vectors around a few
of the deepest `canals', superimposed on gray-scale plots of $P$, in
two frequency bands. The area shown is indicated in
Fig.~\ref{f-pol349}. The polarization vectors on either side of the
`canals' are quite close to perpendicular, demonstrating that the
`canals' are produced by beam depolarization. This perpendicularity
applies to all `canals', irrespective of frequency band and is very
convincing, especially because everywhere else the polarization
vectors vary quite smoothly (if significantly!).

Beam depolarization creates `canals' that are one beam wide, which is
exactly what we observe. This implies that the 90$^{\circ}$ `jump'
must occur on angular scales smaller than the beamwidth. 
%We have made higher resolution maps to verify this. 
At $\sim 2^{\prime}$ resolution (about twice that in
Fig.~\ref{f-phi349}), the `canals' indeed seem unresolved, but the
decrease in S/N-ratio precludes conclusions on even smaller scales
(the original data have $0.8^{\prime}$ resolution).

Additional evidence that the `canals' are due to beam depolarization
is statistical. We defined `canal-like' points from the observed
values of $P$, as follows. For each point in the mosaic we compared
the observed value of $P$ with the $P$-values in pairs of two
diametrically opposed neighbouring (adjacent) points. If the value of
$P$ in the central point was less than a certain small fraction of the
values in {\em both} comparison points, the point was defined
`canal-like'. This definition mimics the visual detection
`algorithm'.

\begin{figure}
\vbox{}
\psfig{file=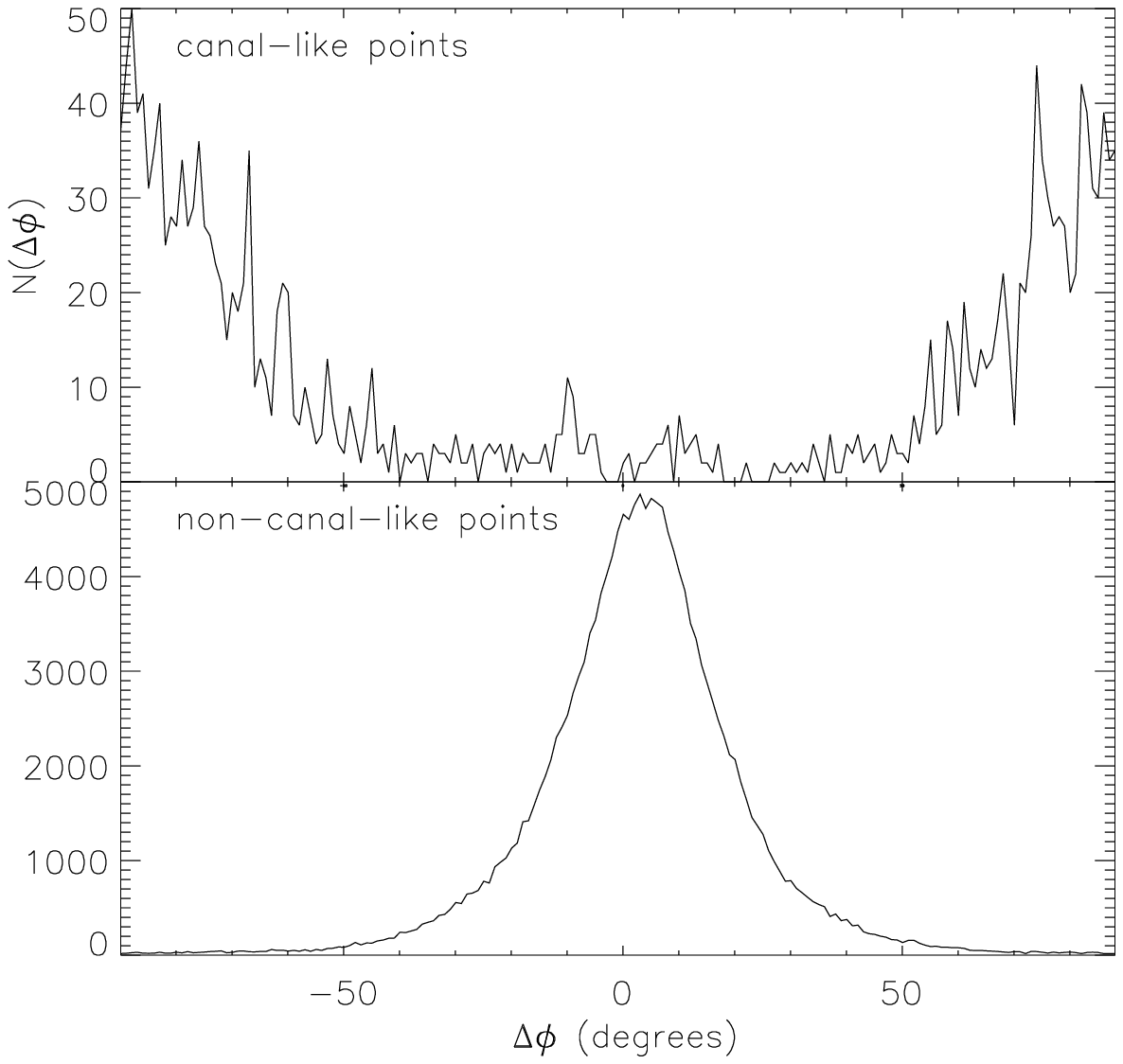,width=8.5cm}
\caption[]{Top panel: Polarization angle difference $\Delta\phi_{\rm pol}$ 
           between two points on opposite sides of a `canal-like'
           point (see text for definition). A clear preference for
           $\Delta\phi_{\rm pol} \approx 90^{\circ}$ across `canals'
           is visible.  \\ Bottom panel: $\Delta\phi_{\rm pol}$
           between two points on opposite sides of non-`canal-like'
           points (see text for definition). For the non-`canal-like'
           points , $\langle\Delta\phi_{\rm pol}\rangle \approx
           0^{\circ}$, rather than $90^{\circ}$. }
\label{f-delphi_c}
\end{figure}

In the top panel of Fig.~\ref{f-delphi_c} we show the distribution of
the difference between the $\phi_{\rm pol}$'s in the two adjacent
points that define the `canal-like' points, for a $P$-threshold of
30\%. The $\Delta\phi_{\rm pol}$-distribution peaks at 90$^{\circ}$,
fully consistent with the beam depolarization hypothesis. This
conclusion is reinforced by a comparison with the distribution of
$\Delta\phi_{\rm pol}$ (again for diametrically opposed adjacent
neighbours) of all points for which $P$ is between 1.0 and 2.0 times
larger than both $P$-values in the two diametrically opposed
neighbouring points, which is shown in the bottom panel of the same
figure.

Similar `canals' were noted by Uyaniker et al.\ (1999) and Duncan et
al.\ (1998), who also invoked beam depolarization. Yet,
Fig.~\ref{f-delphi_c} is the first quantitative proof for this
explanation.

\section{The cause of the `jumps' in polarization angle}
\label{s-rotation}

Two processes can cause jumps in polarization angle $\phi_{\rm pol}$
across the `canals': a sudden change in RM across the `canals', and a
jump in intrinsic $\phi_{\rm pol}$ of the emission incident on the
Faraday screen.  A large change in intrinsic $\phi_{\rm pol}$ implies
a change in magnetic field direction and is therefore quite difficult
to understand in view of the absence of structure in total intensity
$I$ at the more than 2\% level (see Sect.~\ref{s-polint}). On the
other hand, variations in the RM of the Faraday screen would seem to
be quite natural, if not unavoidable.

Discontinuities in RM must play an important r\^ole in producing the
`canals', because the `canals', although similar in adjacent frequency
bands, generally do not occur in all bands, and certainly are not
identical in the different bands (see Fig.~\ref{f-phi349}).  This
indicates that the jumps in $\phi_{\rm pol}$ are mainly due to changes
in RM.  However, the question is if the jumps in $\phi_{\rm pol}$ are
indeed accompanied by jumps in RM {\em of the right magnitude} so that
$\Delta\phi_{\rm pol} = 90^{\circ}$ is produced at the frequency where
the `canal' is best visible.

In principle, the determination of RM only involves a simple linear
fit of the polarization angles in the five frequency bands (at 341,
349, 355, 360 and 375 MHz) vs.\ $\lambda^2$, but in practice several
complications may arise. First, the observed values of $\phi_{\rm
pol}$ may be biased due to imaging effects (like off-sets) in the
Stokes $Q$- and $U$-maps from which $\phi_{\rm pol}$ is derived (cf.\
Wieringa et al.\ 1993). Our data indicate that, in the maps of this
region of sky, such off-sets are quite small, so that the bias in the
$\phi_{\rm pol}$ values is small.  Second, it is not obvious that the
assumption of {\it pure} Faraday rotation ($\phi(\lambda) \propto
\lambda^2$) is supported by the data (see Haverkorn et al.\ 2000).

\begin{figure}
\vbox{}
\psfig{file=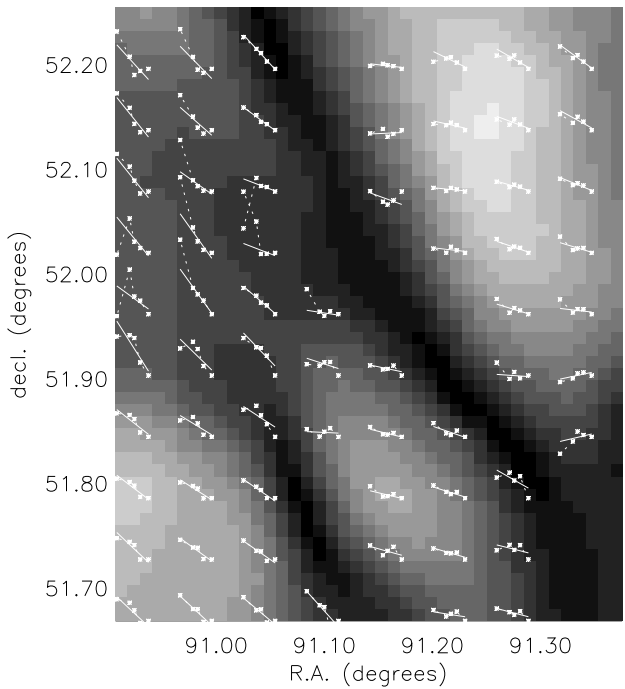,width=8.5cm}
\caption[]{Polarized intensity $P$ at 349~MHz in the area indicated by the 
           box in Fig.~\ref{f-phi349}. Overlaid are small plots of
           $\phi_{\rm pol}(\lambda)$ vs.\ $\lambda^2$ for independent
           points, with a linear fit through the data. RM's range from
           $\sim$ --7 rad m$^{-2}$ in the upper left corner to --1.5
           rad m$^{-2}$ in the `island' between the two `canals'
           (below centre).}
\label{f-rotation}
\end{figure}
 
In Fig.~\ref{f-rotation} we show an array of plots of
$\phi_{\rm pol}(\lambda)$ vs.\ $\lambda^2$ for independent beams in the
small region (indicated in Fig.~\ref{f-phi349}) that contains two
clear `canals'. As can be seen, a direct determination of $\Delta$RM
across the `canals' is not at all trivial. Without knowing the
position of the `canals', one probably would have some trouble to find
the `canals' from discontinuities in RM distribution alone, due to the
uncertainties in the RM-estimates, which sometimes are considerable.
On the other hand, if one knows where the canals are one can identify
some related `jumps' in RM.

From the present data, it seems quite likely that the `canals' are
primarily due to quite abrupt and relatively large changes of RM, with
$\Delta$RM/RM ranging from $\sim$ 0.3 to more than 1 (at least in this
region of sky). Note that in this region the RMs are in the range
from --10 to +10 rad m$^{-2}$ (also confirmed by several polarized
extragalactic radio sources in these same observations). However, a
more robust conclusion about the relation between $\Delta\phi_{\rm pol}$
and $\Delta$RM requires a detailed analysis of more, and more
sensitive data, and a careful error analysis.

\section{Implications for the structure of the local ISM}
\label{s-discuss}

Because we have not yet reached a quantitative conclusion about the
suspected correlation between $\Delta\phi_{\rm pol}$ and $\Delta$RM,
it is not possible to give a full discussion of the implications that
these polarization data have for the small-scale structure of the warm
ISM. However, the data discussed here show the great promise that
high-resolution, multi-band polarization data hold for the study of
the ISM, especially on small scales where pulsars and extragalactic
radio sources cannot give much information.  

Fortunately, more and more sensitive radio polarization data (in
different regions of sky) are forthcoming. In addition, information
must be obtained about the electron density in the warm ISM on the
relevant scales (e.g.\ through H$\alpha$ measurements), as well as on
the other components in the ISM (like e.g.\ the HI).

While we fully realize the preliminary nature of the conclusions
presented, we feel justified to speculate somewhat on the possible
implications of the `canals'. Structure in RM reflects structure in
%the parallel component of the magnetic field and/or the electron
%density 
$B_{\|}$ and/or $n_{\rm e}$ in the ISM. However, as the RM is
an integral over the entire line of sight, the large $\Delta$RM/RM
values that are implied by our observations may give a very specific
message. In particular, we consider it unlikely that the large
$\Delta$RM/RM values are produced mainly by variations in electron
density. Instead, they may be indicating a turbulent ISM with varying
(reversing) magnetic field structures, as modeled in recent MHD
simulations (see e.g.\ Mac~Low \& Ossenkopf 2000; V\'azquez-Semadeni
\& Passot 1999).

\begin{acknowledgements}
The Westerbork Synthesis Radio Telescope is operated by the
Netherlands Foundation for Research in Astronomy (NFRA) with financial
support from the Netherlands Organization for scientific research
(NWO). This work is supported by NWO grant 614-21-006.
\end{acknowledgements}

\vfill 

\end{document}